\documentclass[conference]{IEEEtran}
\IEEEoverridecommandlockouts
\usepackage{cite}
\usepackage{amsmath,amssymb,amsfonts}
\usepackage{algorithmic}
\usepackage{graphicx}
\usepackage{textcomp}
\usepackage{xcolor}
\usepackage{booktabs}
\usepackage{threeparttable}
\usepackage{framed}
\usepackage{adjustbox}
\usepackage{makecell}
\usepackage{dcolumn}
\usepackage{balance}
\usepackage{enumerate}
\usepackage{url}
\usepackage{latexsym}
\usepackage[inline]{enumitem}
\newcolumntype{d}{D{.}{.}{-1}}
\def\BibTeX{{\rm B\kern-.05em{\sc i\kern-.025em b}\kern-.08em
    T\kern-.1667em\lower.7ex\hbox{E}\kern-.125emX}}
\begin{document}

\title{Ideology in Open Source Development\\
}

\author{\IEEEauthorblockN{Yang Yue\IEEEauthorrefmark{1}, Xiaoran Yu\IEEEauthorrefmark{2}, Xinyi You\IEEEauthorrefmark{2}, Yi Wang\IEEEauthorrefmark{2}, David Redmiles\IEEEauthorrefmark{1}}
\IEEEauthorblockA{
% \textit{dept. name of organization (of Aff.)} \\
\textit{\IEEEauthorrefmark{1}University of California, Irvine, Irvine, USA}\\
\textit{\IEEEauthorrefmark{2}Beijing University of Posts and Telecommunications, Beijing, China}\\
\IEEEauthorrefmark{1}\{y.yue, dfredmil\}@uci.edu, \IEEEauthorrefmark{2}\{2020211759, 2020211758, yiwang\}@bupt.edu.cn, %\IEEEauthorrefmark{1}dfredmil@uci.edu
}

}

\maketitle

\begin{abstract}
Open source development, to a great extent, is a type of social movement in which shared ideologies play critical roles. For participants of open source development, ideology determines how they make sense of things, shapes their thoughts, actions, and interactions, enables rich social dynamics in their projects and communities, and hereby realizes profound impacts at both individual and organizational levels. While software engineering researchers have been increasingly recognizing ideology's importance in open source development, the notion of ``ideology'' has shown significant ambiguity and vagueness, and resulted in theoretical and empirical confusion. In this article, we first examine the historical development of ideology's conceptualization, and its theories in multiple disciplines. Then, we review the extant software engineering literature related to ideology. We further argue the imperatives of developing an empirical theory of ideology in open source development, and propose a research agenda for developing such a theory. How such a theory could be applied is also discussed.    
\end{abstract}


\section{Introduction}
Since the 1990s, open source software development has gained momentum in the world of software development, and brings a radical change in the software landscape \cite{Fitzgerald2006OSS}. In contrast to traditional commercial software, which is usually developed by a group of designated developers in a commercial company, and delivered to clients, open source software is a collective product by a group of developers who voluntarily contribute their time and effort, the source code is usually publicly open, and everyone could freely use and distribute with particular licenses.

Open source development is not merely a different software development paradigm, it is much more than that \cite{o1999lessons}. To a great extent, it is a social movement having profound implications beyond the technical realm \cite{diani1992the}. Researchers in the social sciences have long acknowledged the critical role of ideology in driving social movements \cite{mccright2008the,valocchi1996the,zald2000ideologically}. For instance, Leveille \cite{leveille2017searching} offered a detailed case study of how ideology shaped the dynamic of the Occupy Wall Street (OWS) movement between the fall of 2011 and 2012. Moreover, sociologists often argue that the most significant feature of the process of a social movement is the diffusion and evolution of the underlying ideology \cite{rochon1998culture}. Thus, by analogizing to many other social movements, it is fair to argue that ideology plays a vital role in driving the open source movement's development through ideologically coherent social forces. While in this process, its connotation and denotation also evolve. 

At the micro-level, ideology matters to individual's choices and decisions \cite{chen2006relation}. We have long known that a substantial amount of people's decisions, to name a few, voting for which politician \cite{jacoby2009vote}, buying environmentally friendly products \cite{johnston2008consumer}, marrying someone sharing similar religious ideology \cite{Sigalow2012Marriage}, are determined by ideologies. Also, in an organization, ideology determines individuals' participation\cite{Daniel2018ideology}. Imagine a scenario where newcomers are considering which OSS project to join. They might not only choose the project with the skill set they are familiar with, but also prefer the project with a community sharing the same interests as themselves, e.g., building an inclusive project community, encouraging everyone to share and study the code, which is the ideology of the project community.

Given the importance of ideology in open source development at the macro-, meso-, and micro-level, it is striking to find that the ideology(ies) in open source development has not yet received much attention in the software engineering (SE) literature (see our brief literature review in Section V). One of the main reasons for the lack of ideology research in SE may be the inherited vagueness of the conceptualization of ``ideology.'' In this article, we attempt to clarify the concept of ideology by summarizing its historical development, revealing its plural meanings and approaches, examining the extant SE literature related to it, and finally connecting it with open source development. Thus, we argue the necessity of developing an empirical theory around ideology in the context of open source development. A research approach is proposed accordingly. We also blueprint some future implications of the empirical theory to be developed.

The rest of this article proceeds as follows. Section II presents a brief history of the concept of ideology. Section III and IV overview the theories of ideology and organizational ideology in social sciences. Section V offers a literature review of ideology-related work in software engineering. Section VI presents our approach towards the empirical theory of OSS ideology, and the implications are discussed in Section VII. Section VIII concludes the article.

\section{A Brief History of Ideology}

\subsection{Origin of the Concept of ``Ideology''}

The term ``ideology'' has a circuitous history \cite{Thompson1990}, and it has been used by different people with various meanings for over two centuries, in the areas of social science, political analysis, etc. French philosopher Destutt de Tracy first used the term ``ideology'' in 1796. He proposed a new project of systematic analysis of ideas and sensations and argued that, instead of the things themselves, we could only know the ideas formed by our sensation of them, which is our scientific knowledge. Thus, ideology at that time was a ``science of ideas'', originally a positive, useful and rigorous concept\cite{KENNEDY1982}, which de Tracy considered as ``the pre-eminent science that would facilitate progress in human affairs'' \cite{Thompson1990}, corresponding with the spirit of the Enlightenment. Later, he extended the scope of ideology to the social and political areas, analyzing the experience, feeling, thinking, etc. The focus of ideology also shifted to the ideas themselves, instead of the science of ideas at the beginning, and ideology became gradually abstract and illusory ideas \cite{Thompson1990}.

Napoleon Bonaparte criticized the idea of ideology in the political context at the time. He argued that ideology intended to determine political principles based on abstract reasoning, which would encourage rebellion actually. Thus, the concept of ideology was viewed as a political threat from republicanism, and he also blamed it for the collapse of his empire. With de Tracy's exploration and Napoleon's opposition, the term ``ideology'' emerged and started its circuitous history and fuzzy life.

\subsection{Marx's Contribution to Ideology}

Karl Marx played a critical and unique role in the development of the concept of ideology \cite{VanDijk1998}. Marx studied de Tracy's work and Napoleon's negative view on ideology when he was in Paris. Initially, he followed Napoleon's attack on the ideology, and utilized it to criticize the ideas of the ``Young Hegelians.'' In \textit{The German Ideology}, Marx argued that their views are ``ideological'', and they overestimated the value and role of ideas in history and social life \cite{Marx1845}. Furthermore, Marx, along with Engels, also considered the concept of ideology in a more general perspective, analyzing the production and the relations between classes. Thus ideology serves a systematic role in Marx's theoretical framework, and refers to ``the ideas of the ruling class''. In his well-known work \textit{Manifesto of the Communist Party}, Marx explained the socio-economic transformation, unmasked the ideological forms of consciousness of the dominant class in the society, and anticipated the victory of proletariat ideology, with the demise of bourgeois ideology. The concept of ideology had a giant leap and advanced to a higher level with the unique contribution of Marx, which also influences the development of it in the modern era.

\subsection{Modern and Post-modern Development}

Since Marx, the concept and discussion of ideology are widely viewed in terms of its roots in individual and group interests. It has been used in many fields of social science, and has been developed with rich interpretations, not just limited to Marxism. As Marx took over the concept of ideology from Napoleon, it had a negative sense in most of his work. One of the major tendencies in modern and post-modern development, however, was neutralizing the concept of ideology \cite{Thompson1990}.

The neutralization was first started by some Marxists when Marxist social movements in Europe had challenges in particular social-historical circumstances. Lenin faced the polarized political situation in Russia, and he intended to combat the bourgeois ideology and realize the revolution, thus he elaborated ``socialist ideology'', and Luka\'cs, facing a similar situation in the working-class movement, emphasized the importance of ``proletarian ideology'', which both helped the proletarian class understand their situation and class consciousness \cite{Lukacs1972}. In both Lenin and Luka\'cs's work, the concept of ideology is generalized--the ideas to express the interests of the major classes, not associating with a strong negative sense \cite{Thompson1990}. 

Besides Marxists, other scholars also contributed to the neutralization process. Different from Marxists, Karl Mannheim did not view ideology as a weapon of a class for the revolution; he considered it as a method of research in social and intellectual history, which is similar to de Tracy's idea, referring to ``the sociology of knowledge'' \cite{Thompson1990}. He intended to use his new approach to analyze all the factors that influence thought or idea in the social context, which was called the general formation of the total conception of ideology \cite{Thompson1990}. In ``La production de l'idéologie dominante'', Pierre Bourdieu and Luc Boltanski offer critical insights into the principal characteristics and functions of ideologies in advanced capitalist societies \cite{bourdieu1976production}. For example, they argue that ideology's primary function is to ``orient an action or a set of action'', which indicates the shift from ideology's theoretical dimensions to practical dimensions in real-life situations. 

With the circuitous history of development, the term ``ideology'' has been embodied with rich interpretations, and the concept of ideology tends to be general and abstract, also vague and ambiguous at times. However, it becomes clear that \emph{the contemporary uses of ideology have already deviated from the original emphasis on conflicts and tensions among social classes, and come to have considerable overlap with its (social) psychological impacts on individual and groups, and their actions}. We will introduce some representative ideology theories in the next section.

\section{Theories of Ideology}
\subsection{Marxian Theories of Ideology}
Ideology is referred to in various context with multiple conceptions in Marx's work. Thompson summarized them as polemical, epiphenomenal, and latent conceptions in general \cite{Thompson1990}. First, the polemical conception of ideology was derived when Marx criticized the ``Young Hegelians'', which could be defined as ``a theoretical doctrine and activity which erroneously regards ideas as autonomous and efficacious and which fails to grasp the real conditions and characteristics of social-historical life'' \cite{Thompson1990}. Although this conception is a negative view inherited from Napoleon, Marx developed it beyond the original scope, where the ideas should be emerged based on the practice of activity, instead of autonomy. 

Marx and Engels intended to empower the ideology a more general role in the socio-historical analysis, which is the epiphenomenal conception. This conception of ideology is defined as ``a system of ideas that expresses the interests of the dominant class but represents class relations in an illusory form'' \cite{Thompson1990}. Generally, ideology reflects the ruling class's ideas at the particular historical period, and the relations of different classes in the society perceived by the ruling class. This conception is the result of the examination on the production and the diffusion of ideas during the movement of capitalist. And the Marxism theory proposed in the \textit{Manifesto} is around the epiphenomenal conception, which guides the proletariat with the knowledge and experience to understand the class consciousness, become the revolutionary class, and emerge in the new era.

Furthermore, Thompson also categorized the latent conception in Marx's work, which is ``a system of representations which serves to sustain existing relations of class domination by orientating individuals towards the past rather than the future, or towards images and ideals which conceal class relations and detract from the collective pursuit of social change'' \cite{Thompson1990}. This conception mainly characterizes the phenomena of symbolic life in the society, such as slogans and traditions, which are not explicitly presented in the epiphenomenal conception. Marx analyzed them because these symbols play a significant role in orienting people to specific directions, although they are not abstract enough.

Other Marxists, such as Lenin and Luka\'cs, inherited Marx's theory and also further developed the theory, with ``socialist ideology'', ``proletarian ideology'', etc. However, the conception of ideology remains consistent. In general, ideology could be defined as ``the prevailing ideas of an age'' \cite{VanDijk1998}, which ideas reflect the interests of the dominant class in various aspects, such as laws and literature. Moreover, Marxian theories shed light on the dominant class in the future.

\subsection{Non-Marxian Theories of Ideology}

Non-Marxists' theories of ideology cover a wide spectrum in politics, social science, philosophy, etc. Mannheim's work is representative among them. In his theory, the conception of ideology is ``the interwoven systems of thought and modes of experience which are conditioned by social circumstances and shared by groups of individuals, including the individuals engaged in ideological analysis'' \cite{Thompson1990}, which is the general formation of the total conception of ideology. His theory aims to provide ``a revised view of the whole historical process'' \cite{Mannheim1939}, and ideology is considered as ``the systems of thoughts or ideas'' that are socially situated and collectively shared, and the analysis is the study of how the ideology is influenced by the social and historical circumstances \cite{Thompson1990}. And Mannheim found various social factors that could influence the ideology, such as location and generation of the class \cite{Longhurst1989}. 

However, Mannheim's theory also causes ambiguity. On the one hand, he emphasized the generalization of ideology due to the limitation of earlier theories of ideology and applied his research method in a social and historical context. On the other hand, he tried to avoid using the term ``ideology'' to refer to the sociology of knowledge, and acknowledged that some problems still exist with the general conception of ideology \cite{Mannheim1939}.

\subsection{Two Ways to Approach ``Ideology''}
In general, there are two ways to approach ``ideology'', i.e., cognitive and social \cite{VanDijk1998}. Regarding the cognitive way, ideology characterizes the \textit{basis of the social representations}, such as ideas and beliefs, terms used in cognitive science. First, ideas and beliefs are the product of thinking, or the objects/processes in the mind \cite{VanDijk1998}; they could be values, opinions, knowledge, norms, etc. We use ideas and beliefs to perceive the objects around us. For example, ``5\textgreater 2'' could be considered as a belief, which is the result of thinking in mathematics, and also becomes our common sense. However, ideas and beliefs are subjective; hence they could be true or false, right or wrong, good or bad \cite{VanDijk1998}. For example, the statement, ``the earth is the center of the solar system,'' is a valid belief, even though it has been proven to be wrong by science. Generally, the cognitive way to approach ``ideology'' is as an individual's ideas or beliefs about particular things.

Besides the cognitive way, the other way to approach ``ideology'' is the social way. On the one hand, ideology emerges within \textit{social groups}, through the interaction and communication among the members of the group, e.g., what kind of people could become a member of the group. When most of the members reach an agreement on particular things, these usually become part of the ideology. On the other hand, ideology is the things that are shared by the whole group in common, also reflects the common interests of the group. For example, increasing the benefit could be one of the common interests shared by the workers in union organization. Therefore, ideology is the product of the collective effort \cite{VanDijk1998}, and could be approached by the social way.

However, the two ways are neither opposite nor separate when approaching ``ideology''. Indeed, cognitive and social ways define ideology at different levels; individuals' social representations (ideas and beliefs) characterize ideology at the micro-level, while the interests of the social groups capture ideology at the meso/macro-level. The individual's ideas and beliefs could be considered as an initial set of ideology; then they interact with others who also have similar ideas and beliefs, later they could reach an agreement and share the common interests as a whole group. Hence, a general concept of ideology is intertwined with both cognitive and social ways, which is ``\textbf{the basis of the social representations shared by members of a group}'' \cite{VanDijk1998}, and in this article, we use this definition in the study of OSS ideology.

\section{Organizational Ideology}

Since the late 1950s, the concept of ``ideology'' started gaining popularity in many social science fields \cite{hartley1983ideology}. One of such field that is particularly relevant to our inquiry is that of organizational behaviors in management studies. This strand of work is usually referred to as ``organizational ideology.'' As a pioneer in this area, Reinhard Bendix \cite{bendix2013work} studied the social relations in four centuries to identify the nature, origins, and consequences of ``managerial ideologies'', an early notion of ``organizational ideology.'' He defined such ideologies as ``\emph{all ideas which are exposed by or for those who seek authority in economic enterprises, and which seek to explain and justify that authority}.'' Obviously, Bendix's definition is strongly influenced by the Marxist traditions in sociology.

Later, management scholars gradually developed the concept of ideology with distinct discrepancies to sociological traditions. In short, the construct of organizational ideology has shifted from political-economic propositions to shared norms, beliefs, and values by an organization's members at both fundamental and operative levels \cite{Weiss1987}. Such a conceptual adaptation formed the basis for organization behavior scholars to derive new understandings and insights, which effectively connected individuals at the micro-level and organizations at the meso-level. They found that the consensual nature of ideologies may bring people together around organizational actions and thus result in certain competitive advantages \cite{alvesson2018organization}. However, the fundamental and operative ideologies may not be logically consistent or even contradict each other. Abravanel \cite{Abravanel1983} uses the FBI as an example, i.e., while its fundamental ideology claims that ``the FBI is on the side of goodness'', its operative ideology allows various misuse of power. This example suggests that the presence of organizational ideologies may yield rich and complex organizational dynamics, making it is hard to reason the relationship between ideology and other organizational constructs, e.g., organizational outcomes.

Under the umbrella of organizational ideology, researchers have used the theoretical lens of ideology to study many aspects of organizations at both individual and organization levels \cite{gibbs2013overcoming,hinings1996values,kabanoff1991equity, maclean2014living}, to name a few, organizational cohesion, performance, structure, innovation, social responsibility, knowledge sharing, individual social action, engagement, and so on. Despite having many informal features, OSS projects/communities are also organizations. Organizational scholars have already been paying attention to ideology in OSS \cite{Stewart2006ideology}. The extant work in organizational ideology would inform OSS researchers in software engineering the complex interactions among multi-facet ideologies, actions, and organizational constructs and provide models for developing micro-/meso-level theories through rigorous empirical methods. Moreover, the missing link between organizations (meso-level) and society (macro-level) asks for researchers to be sensitive and vigilant in theory development.

\section{Ideology-Related Work in Software Engineering}

In this section, we report a brief literature survey about the studies related to ideology in software engineering; both the paper selection process and the literature survey result will be presented in this section.

In this literature survey, we would like to gain insights into three questions as follows:
\begin{itemize}
    \item How is the term ``ideology'' defined in SE literature?
    \item What are the overall trend of ideology-related work in SE literature?
    \item What are the focuses of ideology-related work in SE literature?
    % \item How 
\end{itemize}

\subsection{Paper Selection}
To build an understanding of ideology-related studies in SE, we conduct a literature survey on the papers published over the past five years, i.e., 2016-2020, and the paper selection process is as follows:

\subsubsection{Venue Selection} We select papers from four top-tier conferences and journals in SE, i.e., the International Conference on Software Engineering (ICSE), the ACM Joint European Software Engineering Conference and Symposium on the Foundations of Software Engineering (ESEC/FSE), the IEEE Transactions on Software Engineering (TSE), and the ACM Transactions on Software Engineering and Methodology (TOSEM). Regarding ICSE, we include all papers from the main track, and also include the papers from the Software Engineering in Practice (SEIP) and the Software Engineering in Society (SEIS) tracks, because papers from these tracks are also full research papers and often focus on the human or social aspects in SE, which could be potentially relevant to ideology. For ESEC/FSE, we only consider the papers from the main track. And for both TSE and TOSEM, we include all published papers. In total, we have 1,698 papers published in these venues from 2016 to 2020.

\subsubsection{Paper Screening} The next step is selecting the papers relevant to ideology from the published papers. Considering the ambiguity and vagueness of the concept of ideology, it is impossible to explicitly define some criteria to automatically screen these papers, as some papers might focus on particular elements of ideology without using the term ``ideology'' or mentioning the concept of ideology. Hence, we manually screen all collected papers. Two of the authors independently review and screen the papers as relevant or not relevant to ideology, based on their title, abstract, and keywords. The inter-rater reliability is 0.98 (Cohen's kappa), indicating excellent agreements \cite{irr}. Then they jointly resolve the disagreements and select $119$ papers \footnote{The list of papers is available: \url{tinyurl.com/IdeologyInSE}} relevant to ideology (see Tab. \ref{tab:relevant}). Furthermore, they discuss the scheme to determine the research focus of each paper, which is based on the title and abstract of paper, then code all selected papers together.

\begin{table}[!h]
\centering
\begin{threeparttable}
    \caption{Distribution of the papers relevant to ideology.}
    \label{tab:relevant}
    \begin{tabular}{lrrrrrr}
        \toprule
        \textbf{Venue \& Track} & \textbf{16} & \textbf{17} & \textbf{18} & \textbf{19} & \textbf{20} & \textbf{Total} \\
        
        \midrule
        \emph{ICSE-Main Track} & 4 & 4 & 10 & 7 & 13 & 38 \\
        \emph{ICSE-SEIP} & 2 & 2 & 6 & 1 & 1 & 12 \\
        \emph{ICSE-SEIS} & 4 & 3 & 3 & 4 & 6 & 20 \\
        \emph{ESEC/FSE-Main Track} & 6 & 2 & 4 & 3 & 10 & 25 \\
        \emph{TSE} & 2 & 3 & 0 & 5 & 8 & 18 \\
        \emph{TOSEM} & 0 & 1 & 0 & 2 & 3 & 6 \\
        \textbf{Total} & 18 & 15 & 23 & 22 & 41 & 119 \\
        \bottomrule
    \end{tabular}
\end{threeparttable}
\vspace{-1em}
\end{table}

\begin{figure*}[!h]
    \centering
    \includegraphics[width=\textwidth]{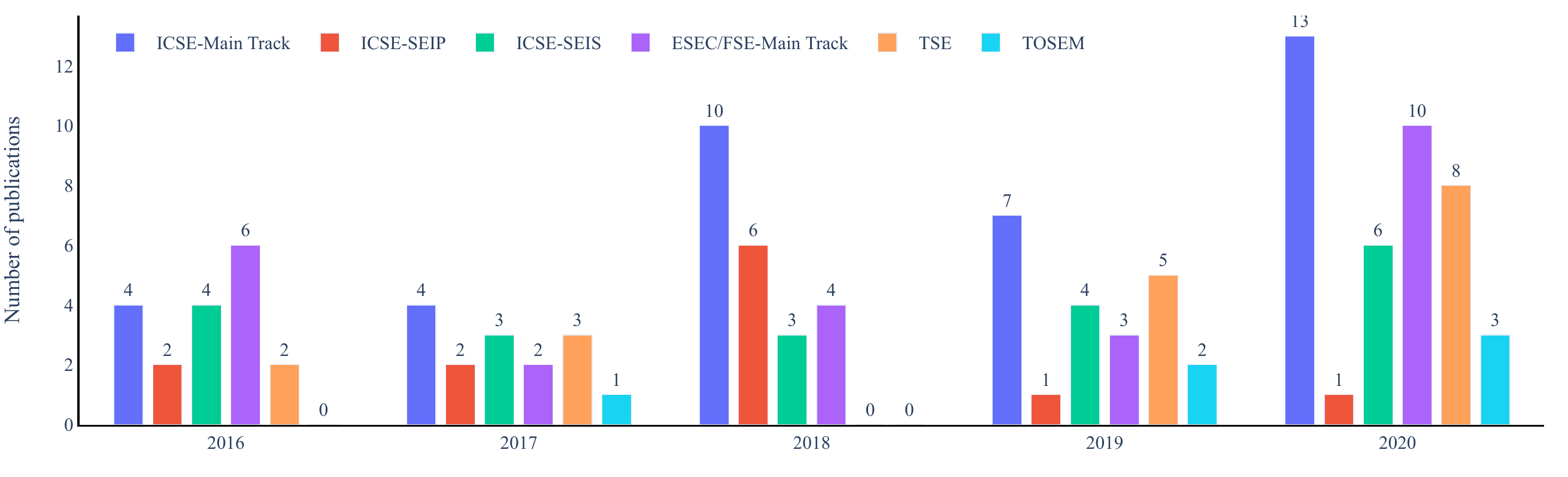}
    \caption{The number of publications in each venue from 2016 to 2020.}
    \label{fig:result}
    \vspace{-1em}
\end{figure*}

\subsection{Results}
Guided by the three main questions, we provide answers to them in the rest of this section.

\subsubsection{Defining Ideology} Among all selected papers, none of them explicitly mentions the notion of ``ideology'', nor defines ideology in SE. However, they touch on some elements of ideology and offer partial knowledge of ideology in their findings.

\subsubsection{General Trend} Over the past five years, the numbers of papers relevant to ideology in SE exhibit an increasing trend, from 18  in 2016 to 41 in 2020 (see Fig. \ref{fig:result}). But in some particular venues/tracks, e.g., ICSE-SEIP, the numbers are slightly decreasing. While the percentages of the papers related to ideology are increasing over the years, they only account for a tiny proportion of SE literature ($\approx 7\%$). Moreover, the percentages in some specific venues are even lower, e.g., only $6.25\%$ papers published in TOSEM are related to ideology.

\subsubsection{Research Focus} 
\begin{figure}[!h]
    \centering
    \includegraphics[width=\columnwidth]{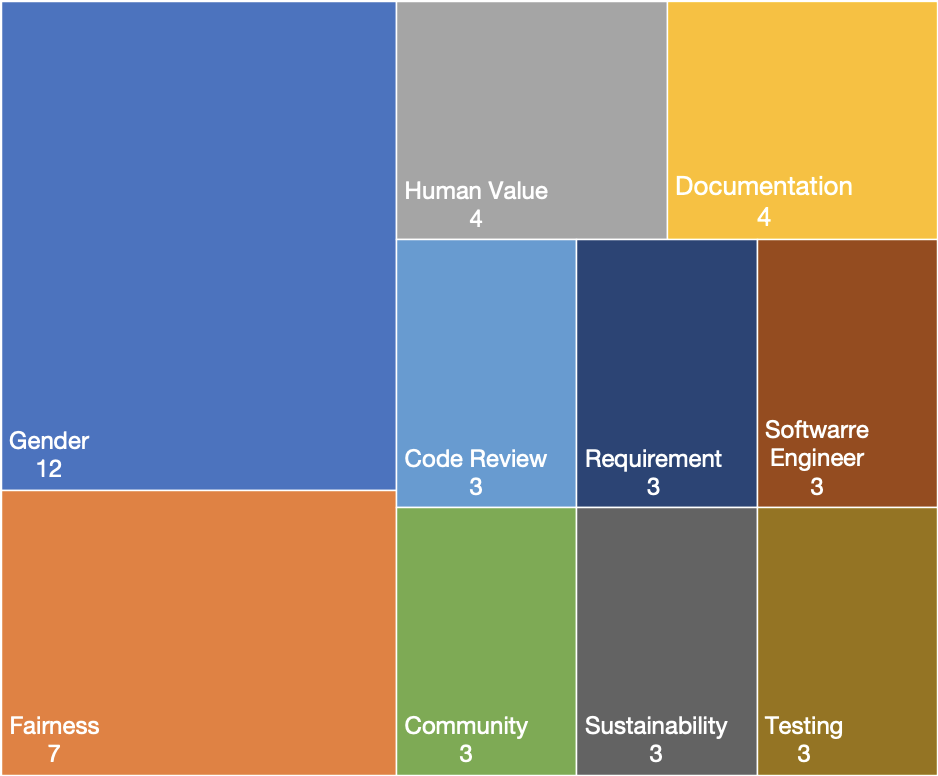}
    \caption{Top 10 research focuses of relevant publications in 2016-2020}
    \label{fig:focus}
    \vspace{-1em}
\end{figure}

The research focuses (determined by the title and abstract) of these papers are diverse, as shown in Fig. \ref{fig:focus}. Generally, we could categorize them into two types: focusing on individual/organization, and focusing on artifacts. Most papers focus on individual/organization (75 of 119). Ideology is a concept touching human or society, which is naturally reflected in individuals or organizations in SE. At the individual level, researchers mainly focus on individual development \cite{Kalliamvakou2018Developer} and work practice \cite{Meyer2017WorkLife, Butler2018WorkPractice}, e.g., the motivation of one-time/episodic contributors \cite{Lee2017OneTime, Barcomb2019EpisodicMotivation, Barcomb2020Episodic}, understanding the older developers \cite{Kopec2018Older} and developers with disability \cite{Armaly2018Disability}. At the organization level, researchers usually study values or beliefs within the community. For example, gender diversity and inclusiveness are among the frequently studied topics in the past five years. Researchers have studied the perception of inclusiveness \cite{Lee2019GenderPerception}, investigated the impacts of gender bias/diversity \cite{Imtiaz2019GenderEffect, Catolino2019GenderSmell}, and explored possible causes \cite{Wang2018Gap} and solutions to improve the status quo \cite{Wang2020ReduceBias}. Besides, some other aspects of team and community are also explored, such as coordination \cite{Herbsleb2016Coordination}, transparency \cite{Valk2018Transparency}, trust \cite{Craggs2019Trust}. Regarding the papers focusing on artifacts, they basically cover every phase in software development, such as requirement and testing, and intend to improve the development practice or embed some values in the artifacts generated in these phases. For example, researchers investigate how requirement practices evolve in startups \cite{Gralha2018StartupRequirement}, explore the ways to embed fairness into ML systems \cite{Aggarwal2019Fairness, Chakraborty2020Fairway}, and improve the accessibility of application \cite{Alshayban2020Accessibility}.

In addition, we also examine if these papers target OSS in their studies, as we are interested in ideology in OSS development. We find almost one-third papers having OSS development as their targets. These papers are related to multiple aspects of OSS with the presence of several elements in ideology. For instance, improving gender diversity, a key element of ideology, could help mitigate some specific unfavorable community smells such as \textit{Black-Cloud} and \textit{Radio Silence} \cite{Catolino2019Smells}. These ``community smells'' are often widespread in OSS development because OSS communities, as social institutions, would inevitably suffer some detrimental organizational and socio-technical issues \cite{Palomba2021Smells}. In this sense, the ideology elements connect to institutional practices and processes, and thereby impact their outputs.  

%For instance, OSS development, as a social activity, would incur so-called ``community smells'', which could affect the outcomes of OSS development . Researchers focus on the relations between gender diversity and community smells, they build statistical models to analyze such relations and reveal that the participation of women could improve particular community smells, such as \textit{Black-Cloud} and \textit{Radio Silence} \cite{Catolino2019Smells}. Their studies provide evidence on how OSS ideology, such as gender diversity, influences OSS projects and communities.

\subsubsection{Summary of Findings}
Our literature survey leads to several findings regarding the extant work related to ideology in SE. First, SE researchers start to pay attention to ideology, and exhibit fast-growing research interests, evidenced by the growing numbers of papers in premier SE venues over the years. However, these numbers are still relatively small, suggesting there are still large gaps remaining unfilled.
%indicating more attention and effort could be utilized in this field.
Second, the research focuses are diverse and cover various aspects of SE (see Fig. \ref{fig:focus}), including the exploration of the elements of ideology and the inquiries if these elements' impacts, with a substantial proportion of papers targeting OSS development. These papers provide rich insights which are beneficial for both software practitioners and researchers. While some elements of ideology have been explored in SE and OSS, they appear as unconnected knowledge fragments, preventing us to form a panoramic view of ideology in OSS. Hence, to 
develop deep understandings of OSS ideology in empirical contexts, a theoretical framework to define, accommodate, organize, and connect its elements is of high necessities. In this article, we anticipate such a theoretical framework of OSS ideology would be expressed as an empirical theory.

In summary, we have several insights derived from the literature survey on ideology-related work in SE:
\begin{itemize}
    \item The extant literature in SE lacks an explicit conception of ``ideology''. They neither use the term ``ideology'', nor present the definition of ideology in SE.
    \item The numbers of related publications are increasing in recent years, indicating growing research interests.
    \item The research focuses are diverse but often unable to connect individuals, organizations, and society, which restricts the prospect of developing a unified theory of ideology in SE.
\end{itemize}

\section{Towards an Empirical Theory of Ideology in Open Source Development}
In this section, we will present the potential of building an empirical theory of OSS ideology, including its definition, framework, and the way leading us to it.

\subsection{Empirical Theory of Ideology In OSS--What Does It Look Like?}

The first step of investigating any social construct is to define it in its context. Regarding the concept of ideology, one of the general definitions is ``the basis of the social representations shared by members of a group''. In the context of open source development, OSS ideology could be defined as \textit{the basis of social representations regarding OSS development shared by the members of OSS community}, which the \textit{social representations} could be values, beliefs, activities, knowledge, or other social/cognitive terms. We intend to keep the definition general, as it could cover every aspect of OSS ideology.

A general conception could include various elements that relevant to ideology as many as possible, however, these elements could be confusing or overwhelming due to lack of a framework to organize them. One of the frameworks employed in the extant interdisciplinary literature \cite{Stewart2006ideology, Daniel2018ideology} is the three-tenet framework, i.e., \textit{beliefs}, \textit{values}, and \textit{norms}, which \textit{beliefs} are understandings of causal relationships, \textit{values} are preferences for some behaviors or outcomes over others, and \textit{norms} are behavioral expectations \cite{Trice1993ideology_theory}. For example, in the context of open source development, ``the source code is freely available'' could be categorized as one of the \textit{beliefs} in OSS ideology, and ``voluntary cooperation is important'' could be considered as one of OSS \textit{values} \cite{Stewart2006ideology}. 

The three-tenet framework is a subtle framework that organizes these detailed elements, but it only considers the \textit{social representations} as \textit{beliefs}, \textit{values}, and \textit{norms}. Besides, the \textit{social representations} could also be activities or relations. In order to establish a comprehensive understanding of OSS ideology, we intend to employ a more general framework proposed by van Dijk \cite{VanDijk1998}, which consists of six aspects:

\begin{itemize}[]
    \item \emph{Membership}: Who are OSS contributors? Where are they from? What do they look like? Who belongs to them? Who can become a member of OSS contributors?
    \item \emph{Activities}: What do OSS contributors do? What is expected of them?
    \item \emph{Goals}: Why do OSS contributors do this? What do they want to realize?
    \item \emph{Values/norms}: What are the main values in OSS development? How do they evaluate themselves and others? What should (not) be done?
    \item \emph{Position and group-relations}: What is their social position? Who are the opponents of OSS development? Who are like them, and who are different?
    \item \emph{Resources}: What are the essential social resources that OSS contributors have or need to have?
\end{itemize}

Based on these aspects and the corresponding questions to investigate, we could establish a comprehensive understanding of OSS ideology, and the next step is how to develop it in the context of open source development.

\subsection{Empirical Theory of Ideology In OSS--How to Develop It?}

\subsubsection{Two-pronged Approach and Its Limitations}
In the extant literature, the ``two-pronged approach'' is utilized to explore the OSS ideology \cite{Stewart2006ideology}. The first one is the narratives of some famous icons in OSS movement, e.g., Richard Stallman, the founder of GNU project and Free Software Foundation, Eric Raymond, founder of Open Source Initiative and an OSS advocate influenced many OSS projects, some of their opinions and views on OSS could be related to OSS ideology. The other one is the narratives from the previous literature that related to beliefs, values, or norms in OSS \cite{Stewart2006ideology}. 

Note that this ``two-pronged approach'' was used before the 2010s, and it yielded an understanding of OSS ideology at that time. However, we cannot simply use this approach to explore OSS ideology with our framework. In the beginning, OSS emerged from ``hacker culture'' \cite{Fitzgerald2006OSS}, the OSS community was a small group of developers, and those famous icons, some of them as initiators, could significantly influence the whole community and movement, which means the views of famous icons basically represented OSS ideology at that time. But OSS has been evolving quickly in recent decades. It becomes a prevalent software development process that attracts a large number of developers to participate due to the availability of several OSS platforms such as GitHub, and the popularity of OSS projects, e.g., Android, Linux. The OSS contributors tend to be more diverse, and even some commercial companies that used to oppose OSS development now start to embrace OSS, e.g., Microsoft and IBM. All those changes make the influence of famous icons limited in OSS communities nowadays. Moreover, it also contradicts the spirit of ideology theory, which ideology should reflect the interests of the majority, instead of a small group of people. Hence, comparing with the famous icons, the narratives of the grassroots, i.e., individual OSS developers, tend to be more important and representative.

On the other hand, based on our literature survey, the number of relevant publications are limited. Some aspects of our framework might not yet have been explored. For instance, the extant literature focuses on diverse topics relevant to OSS, but the elements of ideology investigated tend to be scattered, which failed to connect the micro-, meso- and macro-levels of ideology. Furthermore, most literature lacks the integration of multi-disciplinary literature, ignoring the interdisciplinary nature of ideology. Thus, the narratives from the previous literature might not provide enough knowledge to establish a comprehensive understanding of OSS ideology. 

\subsubsection{Our Approach}
Considering the current development of OSS and the limitation of the ``two-pronged approach'', we plan to employ a grounded theory methodology in developing the proposed empirical theory of ideology in the context of OSS development. The approach would rely on a greatly diverse sample of OSS practitioners and incorporate a richer body of literature from multiple disciplines. Theoretical coding will be used to link the empirical observations with the literature. The approach consists of three major iterative processes: sampling, data collection, and analysis.

We would sample potential participants through a process similar to the Address-Based Sampling (ABS) \cite{Messer2011ABS}. We view the major OSS project hosting platforms as OSS practitioners' ``home addresses''. We set the initial study invitation number as 2000. With a 5-10\% response rate, it could result in about 100-200 potential participants. The number of invitations sent to each platform's practitioners is proportional to the total number of users of the platforms. These potential participants will be screened, selected, and contacted for potential audio/video interviews. This method enables us to reach a diverse sample of OSS practitioners, particularly some grassroots practitioners. Then, their narratives will be collected through semi-structured interviews. We would ask a series of questions related to OSS ideology. They also have the freedom to talk about their opinions and experiences they would like to share. Upon the participants' consent, the interview would be either audio-recorded or noted. We expect each interview will last for about 30-45 minutes. 

As an inductive study, the data analysis for theorizing occurs during and after data collection \cite{Thomas2003Inductive}. Following grounded theory \cite{Corbin2014Basics}, we start the data analysis parallel to data collection, using open coding with the assistance from the Atlas.ti qualitative research software. In this process, transcripts of participants' will be studied to determine the exact meanings of their narratives, and assign an appropriate provisional code to each passage. Multiple researchers will independently code the data, then share, discuss, and resolve the coded transcript and descriptive memos about emerging themes. 

After the open coding phase, we plan to use the constant comparative method \cite{Glaser1965ConstantComparative} to identify the recurring themes in the data. Axial coding will be used to find connections in the categories. In the last step, we will conduct selective coding to identify broader themes and dimensions that would form the emergent theories. This step focuses on an in-depth understanding of individual categories and integrating them into the six aspects framework of ideology defined by van Dijk. Note that we will not restrict ourselves to van Dijk's framework in case there are findings that might not be integrated into it. Instead, we would adapt it to achieve a good fit of the empirical data, as required by grounded theory \cite{glaser2005staying}. Such an organizing scheme enabled systematic thinking about the phenomenon under study. During the above coding process, we take an iterative approach to move back and forth between theory and data by comparing empirical understandings that emerged from the data with the theoretical insights derived from the extant literature and vice versa. In each step, researchers discuss results in details to ensure the reliability of the analytical process and the mutual agreement on results.

Note that the above process is iterative until theoretical saturation. It is possible that more than one wave of interview invitation would be sent out. Through this process, we could establish a comprehensive empirical theory of ideology in open source development.

\section{Applying the Empirical Theory in Future Research and Practices}
A good theory must be applicable in future research and practice. We argue that the aforementioned empirical theory, if established, would have significant implications to the research and practice of OSS development. In this section, we will envision applying the empirical theory in future research and practice by discussing potential theoretical, empirical, and practical implications.  

\subsection{Theoretical Implications}
\subsubsection{The Empirical Theory as A Terminological System}
The empirical theory to be developed would provide a general terminological system for the group of sociotechnical meanings of ``ideology'' in the context of OSS development \cite{cabre1999terminology}. Our literature survey (section V) indicates that ideology-related work in SE is often tangled in terminological fragments of ideology rather than a system. The empirical theory thus helps organize, connect, and synthesize existing literature systematically in a framework \cite{elkin2012terminology}. Moreover, as we emphasize in this article, the meanings of ideology are plural and multidisciplinary by nature. Communicating ``ideology'' among scholars from different disciplines with distinct academic traditions is challenging. For instance,  sociologists may feel it difficult to accept how management scholars define and handle ideology \cite{Stewart2006ideology, Daniel2018ideology}; software engineering researchers might disagree with media researchers on associating ideology with cultural icons \cite{strodthoff1985media}. Under these circumstances, the empirical theory could maximize the pragmatic discourses and interpretations of contextualized personal and group experiences with ideology while avoiding metaphysical arguments around them, facilitating multidisciplinary academic exchanges. Such a terminological system would allow the discovery of analogy or relation between two academic fields, which leads to each helping the other's progress, repeatedly proven in the modern history of natural and social sciences \cite{abramo2018effect}.       

\subsubsection{The Empirical Theory as A Historical Artifact}
All empirical theories related to social phenomena become ``outdated'' immediately after their establishment. They could only reflect a small proportion of social reality in a specific snapshot of history. However, as a historical artifact, such an empirical theory would serve as a referential point for future theoretical development of ideology in OSS development, which gives us a more comprehensive and dynamic picture of the development and evolution of the open source movement, largely driven by the evolution of its ideologies. Being a historical artifact also means that the empirical theory is by default a subject of deconstruction or critical analysis \cite{caputo1997deconstruction}. Ideal concepts have the natural tendency to be irreducibly complex, unstable, or impossible to determine \cite{derrida1978writing}. Therefore, they must be critically analyzed\footnote{See an example in Chapter 28 of van Dijk (1998) \cite{VanDijk1998}.} by the Constructive Other and open to new meanings as well as refutations \cite{miller2008otherness}. Such an openness ensures the empirical theory to be ethical by exposing and publicizing the historical, social, economic, political, and psychological constraints of the theory, not to mention our subjectivity. 

\subsubsection{The Empirical Theory as A Methodological Shift}
As we noted in section VI.B.2, the empirical theory will be developed with the inductive reasoning approach (grounded theory) with a broader, more diverse sample of OSS practitioners and body of literature far beyond the conventional two-pronged approach. Such a methodological shift is not arbitrary. It is based on the critical changes of the OSS movement in the last couple of decades. OSS development has penetrated almost every aspect of our society, and thus no longer a hacker culture creation \cite{castells2002internet, Fitzgerald2006OSS}. With millions of participants around the world, its ideology must be defined and practiced by such a diverse population rather than a few famous OSS icons. The OSS grassroots, whose opinions and thoughts are often neglected, shall be involved in creating the proposed empirical theory.    

\subsection{Empirical Implications}
The empirical theory to be developed would inform rich opportunities for future empirical work. 

\subsubsection{Understanding Individual's Decisions and Actions}
Weiss \& Miller \cite{Weiss1987} argue: ``the issue of why individuals hold and promote the belief, attitude, and values they do--that is the question of the origin of ideas--is a fundamental one for those interested in how human consciousness affects human social action'' and rigorous empirical studies ``hold the promise of expanding our understanding of the crucial relationships'' between ideology and actions. The empirical theory to be developed serves as a pivot for such empirical studies. We are going to give a couple of concrete examples.

Let us first have a look at newcomers of OSS projects. Ideological beliefs, attitudes, and values have long been suspected of playing a significant role in newcomers' decisions and actions in the process of participating in OSS development \cite{10.1145/3239235.3267427}. For instance, in deciding which projects to join, some elements of ideology may become key concerns. Newcomers whose ideology leans to the progressive side may not want to join a project discriminating against underrepresented groups. However, without theoretical knowledge such as the empirical theory to be developed, researchers may not be able to systematically answer questions such as which are the specific aspects of ideology related to such decisions, and how such aspects interact with other factors in the decision process? Then, let us shift our focus to the development process. It is now common for an OSS project to introduce company sponsors \cite{wagstrom2009vertical}. In this process, it is likely to incur some changes in the project's ideology, and thereby lead to ideological mismatches/conflicts \cite{kernberg1998ideology} between individual members and the project. The empirical theory could help researchers study such mismatches/conflicts to frame their scope, guide their data collection and analysis by enhancing their theoretical sensitivity and knowledge base.

\subsubsection{Understanding Collective Dynamics at Organizational Level}

Ideology reflects the common interests shared by an organization \cite{VanDijk1998}; in turn, individuals' actions impacted by their ideologies also shape the organization's collective dynamics. With an empirical theory, we could gain novel and deep insights into OSS projects' collective dynamics and reasoning about how ideology interacts with many organizational constructs. 
For instance, one of the long-lasting knowledge gaps in OSS research is how ideology could dynamically influence a project's communication structure \cite{boutyline2017social}. With a comprehensive empirical theory of ideology, researchers may be able to locate specific ideological elements relevant to the project's communication structure without missing some essential ones and precisely estimate each element's impact. Moreover, given the pervasive accessibility of OSS project data on the Internet, the empirical theory could be integrated with data science techniques in mining software repositories, thus realizing the mutual reinforcement between theory and data, especially when studying
and supporting the complex dynamic socio-technical systems such as OSS projects.

%Especially we have various OSS projects hosted online nowadays. Even though they are all OSS projects, there exist differences across projects teams. For example, some projects have a large number of contributors, while some projects maintain a small group of contributors since the beginning. The diversity and inclusiveness of the project team, which is also a major concerns within OSS community, is influenced by their perception of OSS ideology. Moreover, regarding the software development activities, some project teams utilize pull-request model, some project teams release new versions every month. The various choices on development models and development processes, and the outcomes are also influenced by the perception of OSS ideology in project teams. Therefore, the empirical theory of OSS ideology could help us better understand these collective dynamics at organization level, such as the factors to improve the team diversity, the factors to choose particular development techniques and processes, how the particular team dynamics reflect the perception of OSS ideology, etc.

\subsection{Practical Implications}
The practical values of the proposed empirical theory are also multifold. 

First, with the empirical theory of OSS ideology to be developed, developing a full-scale measurement system for OSS ideology becomes realistic. Such a measurement system could be implemented in questionnaires, which might be used by practitioners to perform evaluations on themselves and projects. Such evaluations on the basis of explicit empirical theory would help people make informed decisions based on systematic information, and reduce the chances of being biased by some fragile information. In a typical use case, newcomers may use the measurement to evaluate the matches between their own ideology with the ideology of the project they plan to join, thus avoiding the pitfalls resulting from ideology misfit\cite{Daniel2018ideology}. Besides, automated ideology extraction techniques may also be invented under the empirical theory's framework, similar to what researchers have accomplished in automatically extracting many social psychology constructs such as personality, personal values, and so on \cite{vinciarelli2014survey}. Such techniques, if realized, could further reduce practitioners' decision burdens.      

On the other hand, an empirical theory would also help a project better communicate its ideology with potential contributors and the public. Doing so would help them attract more qualified potential contributors sharing a similar ideology. These potential contributors are likely to be motivated by the shared project goals. With such contributors' participation, the project team's cohesion may be improved; cooperation may be more proactive; and turnover may be lowered. All these would contribute to better project outcomes. In return, individuals' personal well-being, e.g., satisfaction, also may be higher than those trapped in an less well-fitting project \cite{francca2018motivation}.

At the societal scale, many social values and truths in ideology, such as fairness, justice, and freedom, are cherished by all human beings. As a social movement, OSS development is relevant to OSS practitioners and almost everyone in our society, with such social values and truths. The empirical theory of OSS ideology could help us recognize and share these social values and truths across the whole OSS community, e.g., concerns about inclusiveness or individual developer's freedom. An empirical theory could also help the public audit whether open source projects and their products uphold such social values and truths. For example, the public could check if a project implements mechanisms and practices for creating harassment-free community or if an OSS project addresses accessibility issues for people with disabilities.

\section{Concluding Remarks}
There has been increasing interest in SE research related to ideology or some elements thereof. The OSS movement, which is largely driven by ideology, serves as a perfect candidate for ideology-related theoretical and empirical inquiries. Thus, the ideology lens could provide rich insights about OSS at individual, organizational, and societal levels. This article argues the necessity of establishing an empirical theory of OSS ideology and presents an approach for achieving such a goal. We start from a brief history of ideology and its theories in social science. Then we report a brief literature survey about the ideology-related work in SE. Even though SE researchers have built a body of knowledge about ideology in SE, some issues, e.g., lack of explicit definition of ideology, still exist, which are obstacles to establishing comprehensive knowledge. Therefore, we propose an agenda for developing an empirical theory of ideology in OSS development. We first define the contextualized concept of ideology and its general framework in OSS development. Then we introduce our approach of theory development using grounded theory. We discuss various implications when applying the empirical theory of ideology in future research from theoretical, empirical, and practical perspectives. Our work sheds some light on research into the ideology of OSS development, which would potentially benefit both researchers and practitioners.

\section*{Acknowledgement}
This work is partially supported by the University of California, Irvine, Donald Bren School of Information and Computer Sciences, and the Fundamental Research Funds for the Central Universities, Beijing University of Posts and Telecommunications. Corresponding Author: Yi Wang.

\bibliographystyle{IEEEtran}
\bibliography{IEEEabrv, reference}

\end{document}